\documentclass[preprint]{iucr}              

     \journalcode{J}              

\usepackage{array}

\newcolumntype{L}[1]{>{\raggedright\let\newline\\\arraybackslash\hspace{0pt}}m{#1}}
\newcolumntype{C}[1]{>{\centering\let\newline\\\arraybackslash\hspace{0pt}}m{#1}}
\newcolumntype{R}[1]{>{\raggedleft\let\newline\\\arraybackslash\hspace{0pt}}m{#1}}

\usepackage{siunitx}

\begin{document}                  



\title{Extending SAXS instrument ranges through addition of a portable, inexpensive USAXS module}


\cauthor[a]{Brian R.}{Pauw}{brian.pauw@bam.de}{fourth argument} 

\author[b]{Andrew J.}{Smith}
\author[b]{Tim}{Snow}
\author[b]{Olga}{Shebanova}
\author[b]{John P.}{Sutter}
\author[c]{Jan}{Ilavsky}
\author[d]{Daniel}{Hermida-Merino}
\author[a]{Glen J.}{Smales}
\author[b]{Nicholas J.}{Terrill}
\author[a]{Andreas F.}{Th\"unemann}
\author[e]{Wim}{Bras}

\aff[a]{Bundesanstalt f\"ur Materialforschung und -pr\"ufung (BAM), 12205 Berlin, \country{Germany}}
\aff[b]{Diamond Light Source Ltd., Diamond House, Harwell Science \& Innovation Campus, Didcot, Oxfordshire, OX11 0DE, \country{United Kingdom}}
\aff[c]{Advanced Photon Source (APS), Argonne National Laboratory, Argonne, IL 60439, USA}
\aff[d]{Netherlands Organization for Scientific Research (NWO), Dutch-Belgian Beamlines at the ESRF, Grenoble, France}
\aff[e]{Chemical Sciences, Oak Ridge National Laboratory, Oak Ridge, TN 37831, US}









\renewcommand{\thefootnote}{\fnsymbol{footnote}}
\maketitle                        

\begin{synopsis}
Supply a synopsis of the paper for inclusion in the Table of Contents.
\end{synopsis}

\begin{abstract}
Ultra-SAXS can enhance the capabilities of existing SAXS/WAXS beamlines and laboratory instruments. A compact Ultra-SAXS module has been developed, which extends the measurable $q$-range with $0.0015 \leq q\mathrm{(nm^{-1})} \leq 0.2$, allowing structural dimensions between $30 \leq D$(nm)$ \leq 4000$ to be probed in addition to the range covered by a high-end SAXS/WAXS instrument. By shifting the module components in and out on their respective motor stages, SAXS/WAXS measurements can be easily and rapidly interleaved with USAXS measurements. 

In this paper, the design considerations, realization and synchrotron findings are presented. Measurements of silica spheres, an alumina membrane, and a porous carbon catalyst are provided as application examples.
\end{abstract}

\section{Introduction}

Small-angle X-ray scattering (SAXS) benefits from expanded measurement ranges, both towards wide angle as well as very small angles. With current laboratory and synchrotron SAXS instruments now able to measure up to four decades in scattering vector $q$\protect{\footnote{Commonly defined as $q = \frac{4\pi}{\lambda}\sin\theta$ with wavelength $\lambda$ and a scattering angle of $2\theta$}} -- thus probing up to four decades in structural details for a given material -- insights are gained not of individual structural components in isolation, but of the complete hierarchical interplay of structures \cite{Narayanan-2018, Smith-2019, Smales-2019, Allen-2008}. This allows for much more comprehensive structure-property relationships to be established, and correlations between the atomic arrangement and nanostructure can be evaluated on a consistent dataset. Indeed, once this Pandora's box of multi-decade-spanning datasets has been opened, all other analyses focusing on a single decade can appear extremely myopic in scope and applicability. 

Extending the higher limit of most point-collimated small-angle X-ray scattering (SAXS) instruments can be done easily by installing a suitable wide-angle X-ray scattering (WAXS) detector. Achieving smaller scattering angles below the typically achievable $q \mathrm{(nm^{-1})} \approx 0.03$, however, requires exponentially longer extensions of the existing equipment and concomitant improvements in collimation. One alternative for funding- and/or geometry-restricted instruments is to add a Bonse-Hart type Ultra-SAXS (USAXS) module instead. 

Bonse-Hart USAXS instruments rely on a high-precision rotation scan of a multi-bounce ``analyzer'' crystal, acting as a narrow-bandwidth angular filter, to pick out the photons scattered by a sample at the slightest of angles from the unscattered beam \cite{Bonse-1966}. This necessitates the primary beam to be of similarly low divergence. To achieve this, an identical multi-bounce crystal is typically employed, which is positioned upstream of both the sample and analyzer crystal. At least the downstream channel-cut crystal needs to be equipped with a fine yaw rotation with a sub-microradian resolution for the USAXS scans. The sample is placed in between the two crystals on a normal stage. Such instruments are very efficient at determining the scattering cross-section close to the direct beam, but become progressively less efficient at larger angles due to the very narrow angular bandwidth of the crystals, and have a limited speed due to their scanning nature, as they can only collect one scattering angle at a time. 

A dedicated high-performance USAXS beamline, such as at the APS (currently located at 9ID), can extend their range to higher angles by swapping out the USAXS analyzer stage with compact SAXS and WAXS modules \cite{Ilavsky-2018}. Due to geometrical restrictions, the accompanying SAXS instrument is not able to reach low in $q$, necessitating the USAXS instrument to continue its measurement into the less efficient regime. We are exploring the opposite arrangement, where a high-performance synchrotron or laboratory SAXS instrument (capable of reaching at least down to $q\mathrm{(nm^{-1})} \approx 0.03$) is extended with a smaller, less exceptional USAXS module. This means that only the range of $0.0015 \leq q\mathrm{(nm^{-1})} \leq 0.03$ has to be covered by the USAXS instrument, beyond which the normal equipment may take over.

\begin{figure}
	\begin{center}
		\includegraphics[width=0.95\textwidth,angle=0]{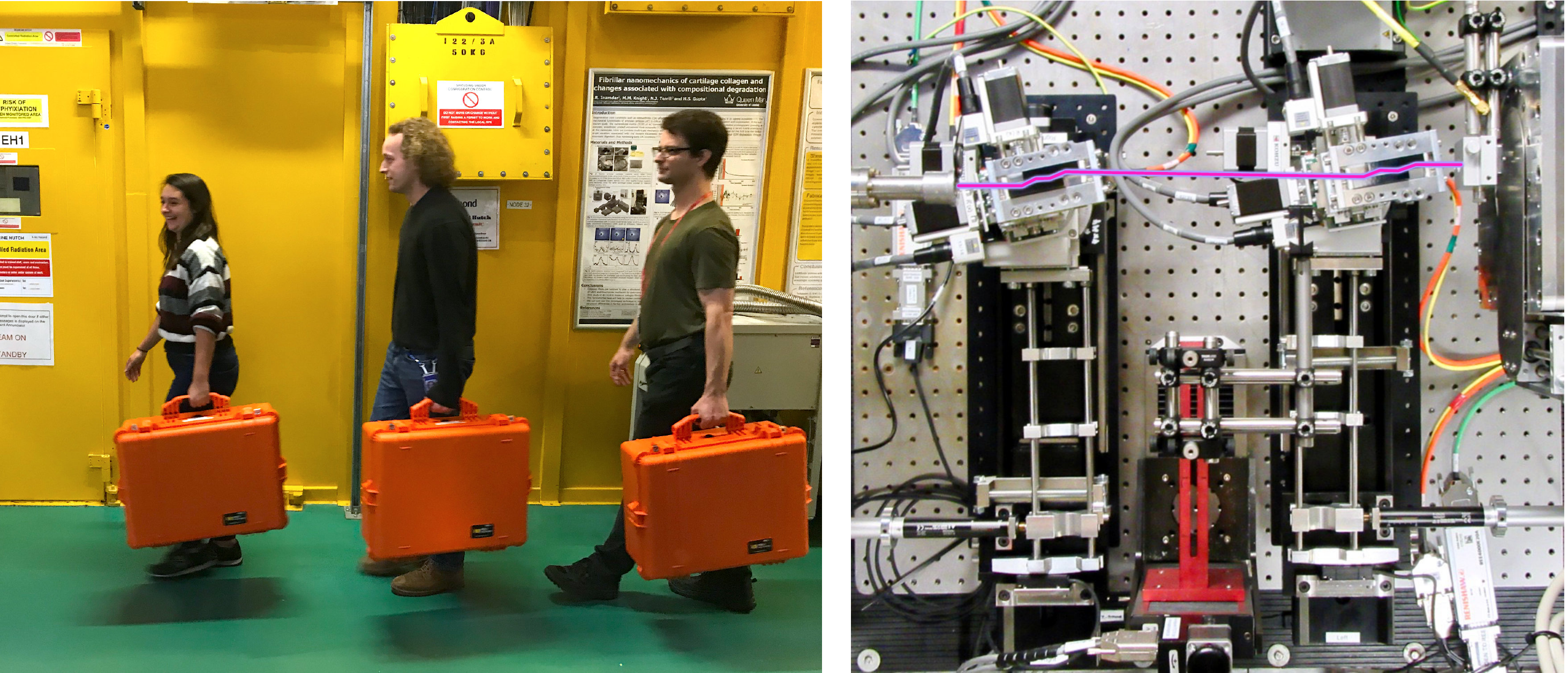}
		\caption{The two crystal stages and the central sample stage can be transported together with a range of accessories in the three flight cases shown on the left-hand side. A top-down view of the USAXS module as installed at the beamline, with the X-ray beam drawn in purple, travelling from left to right through the two channel-cut Si(220) crystals.} \label{fg:CasesAndTopView}
	\end{center}
\end{figure}

In continued discussion with experienced USAXS instrumentalists (and after constructing two prototypes), such a USAXS instrument has now been realized (Figure \ref{fg:CasesAndTopView}). Its construction, costs, and synchrotron performance tests are detailed in this paper. An outlook on its continued evolution and further cost reduction steps is provided thereafter.

\section{Design considerations}

The standard design of a Bonse-Hart USAXS instrument consists of two goniometer stacks (crystal towers), equipped with sufficient motion to position the crystals into the beam. Each stack also has a high-resolution fine-yaw rotation stage, and can be equipped with roll and tilt motions to align the crystals with respect to each other. The two crystals are placed with sufficient (but not excessive) space in between to position a sample. A separate detector, typically a PIN diode detector, is placed close after the downstream crystal to detect the scattered radiation. Depending on the range of rotation of the downstream crystal, its design and the detector aperture, the detector may need to be moved to match the change in beam offset when rotating the downstream crystal.

A portable, ``plug-in'' USAXS module that augments existing high-performance instruments will not necessarily be used for \emph{all} experiments. This significantly alters its central design tenets compared to USAXS instruments built for dedicated USAXS beamlines or laboratories. The requirements for this plug-in USAXS instrument are:
\begin{itemize}
\item{\textbf{cost:}} it has to be sufficiently inexpensive so that it can be an affordable addition to a SAXS beamline or laboratory's repertoire,
\item{\textbf{size:}} it has to be sufficiently compact to fit on existing sample tables and/or vacuum chambers,
\item{\textbf{set-up simplicity:}} its installation and alignment procedure has to be straightforward and fast, requiring no complex changes to the main instrument's configuration, 
\item{\textbf{interleaving capability:}} it should be capable of being moved in and out of the beam in a reproducible and rapid manner, to allow interleaving with SAXS experiments, and
\item{\textbf{universality:}} exotic components are to be avoided, to accelerate integration into the existing beamline control systems
\end{itemize}

\begin{figure}
	\begin{center}
		\includegraphics[width=0.75\textwidth,angle=0]{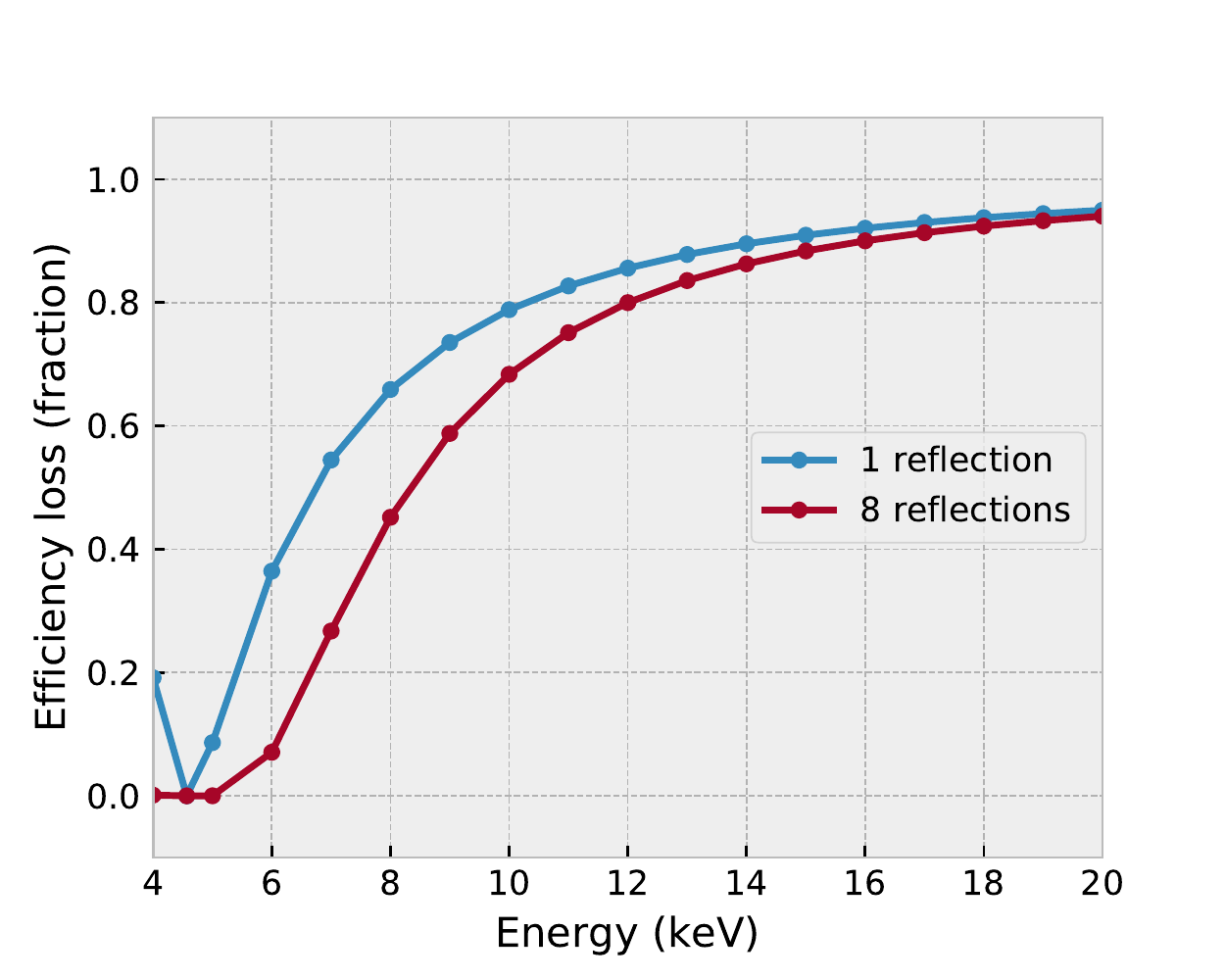}
		\caption{Ratio of the throughput for horizontal vs. vertical reflections for one or eight consecutive Si (220) reflections, with a horizontally polarized beam. A total loss after eight reflections of $\approx 20$ \% is expected at 12 keV, and $\approx 10$ \% at 18 keV.} \label{fg:efficiencies}
	\end{center}
\end{figure}

Building on experience with two earlier prototypes of a simple laboratory USAXS set-up, a new design was made which is compatible with both laboratories and synchrotrons. The added synchrotron compatibility demands more motorization on the two crystal towers of the instrument and the inclusion of encoders on the fine crystal rotations. Similar to the earlier prototypes, the high-precision crystal rotation axes are vertical for improved mechanical stability and reduced complexity. This does imply a loss of crystal reflection efficiency due to their perpendicularity to the synchrotron polarization, but the efficiency losses are acceptable at high energy (see Figure \ref{fg:efficiencies}).  

\begin{figure}
	\begin{center}
		\includegraphics[width=1\textwidth]{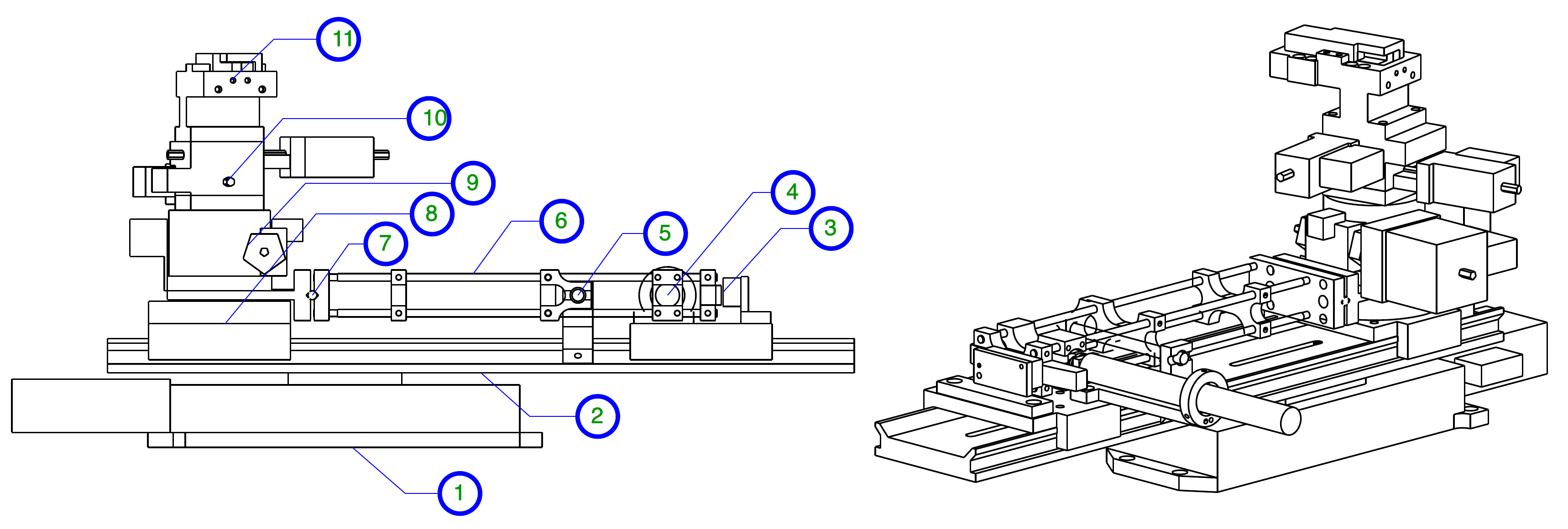}
		\caption{Tower of the crystal rotation. 1: long-travel horizontal translation stage with adapter plates 2: Optical rail 3: encoder readout 4: linear actuator 5: spring return 6: rotation arm cage 7: vertical arm joint 8: rail clamp with cross-roller bearing 9: coarse yaw rotation 10: roll and pitch stage 11: crystal box with crystal} \label{fg:Tower}
	\end{center}
\end{figure}

The earlier prototypes also successfully employed a sine-arm fine rotation design comprised of a cross-roller bearing ring (DIN 620 precision grade P2 in the prototypes), an approximately 300 mm long arm cut from 5 mm thick carbon sheet steel, with a high-resolution linear actuator at its end. For the new design, a more lightweight adjustable arm was developed around a cage system, and a higher-precision cross-roller bearing was selected (USP grade). A linear joint was added for vertical adjustment of the arm, and the arm was tipped with an encoder strip to detect actual arm deflections. The design of the crystal tower assembly is shown in Figure \ref{fg:Tower}. The sample is placed between the two crystals, on a platform large enough to accommodate a range of sample environments. Each of the stages is mounted on a long-travel (100 mm) linear stage, that allow the crystals and sample to be easily moved in and out of the X-ray beam, and can be used to align the tower rotation axis with the beam. 

The cost for the instrument consists of the components listed in Tables \ref{tb:costUDStage}, \ref{tb:costCStage}, and \ref{tb:costAssComponents}. The cost of a monochromatic X-ray source and basic collimation has not been included, nor has the cost of an instrument control system. These have been omitted since the instrument is intended to extend an existing SAXS instrument, and is therefore expected to already include the essentials. 

\section{Experimental}

\subsection{Beamline configuration}

X-ray data were collected at the I22 beamline at the Diamond Light Source, UK (18.0 keV). The SAXS detector was positioned at a distance of 9.40 m from the sample as calibrated using a 100 nm period Si$_3$N$_4$ grating (Silson, UK), giving a usable range of $0.02 \leq q\,\mathrm{(nm^{-1})} \leq 2.5$. The WAXS detector was positioned at a distance of 0.3157 m from the sample as calibrated using a standard CeO$_2$ sample (NIST SRM 674b, Gaithersburg, MD, USA), giving a usable range of $2.14 \leq q\mathrm{(nm^{-1})} \leq 28.9$. Samples were mounted in Thorlabs CFH1-F holders, directly behind a 2 mm lead pinhole.

Simultaneous SAXS/WAXS data were collected in 10 frames of 100ms per sample. Data were corrected and reduced using the DAWN package \cite{Basham-2015, Filik-2017} and standard reduction pipelines \cite{Pauw-2017}. Deconvolution ('desmearing') of the USAXS data for visualization purposes was performed using the IRENA package using a slit-length of $q_s = 0.2$ $\mathrm{nm^{-1}}$. This slit length was estimated by measuring the distance between the sample and the detector, and combining this distance with the detection window dimensions. 

After collection of the USAXS scan, the data is processed using DAWN to correct for the shifting position of $q_0$, the darkcurrent, transmission (calculated using the integral intensities of the sample and background scans), and background. Following calibration, the data is output in slit-smeared intensity vs. $q$, which may be de-smeared for visualization purposes. 

\section{Instrument set-up and alignment}


In general, the alignment procedure must place each crystal channel in the beam with the fine yaw rotation center on the first reflection of each crystal. The crystal channel pitch must be approximately parallel with the beam path, and the pitch and roll of both crystals should match. The crystals are fully aligned one after the other, the upstream crystal motions are no longer adjusted once the second crystal is placed into the beam. Each crystal alignment will first focus on optimizing the beam position along the first crystal reflection channel wall, with the wall parallel to the beam. After this, each crystal is rotated to its diffraction condition, the detector offset to match the beam offset through the crystal, and the coarse and fine yaw rotation further optimized. The exact alignment steps are discussed in detail in the following paragraphs.

A coarse alignment of the crystal orientation is achieved by levelling the crystal housings with a fine spirit level in both roll and pitch directions to $0 \pm 100 \, \mathrm{\mu rad}$. After this alignment is done, the entire table is pitched downwards by 5.2 mrad, to match the downward direction of the primary beam at the beamline. No further optimization of the roll and pitch of the crystals is done, and the respective motorized stages are disconnected to avoid cable strain on the tower. The crystals are roughly placed in the beam path (``align-by-eye''), while simultaneously ensuring that the translation on the long-travel crystal stages have sufficient range for horizontal alignment and for moving the crystals completely out of the beam when necessary.

An automated crystal alignment script aligns the beam with the edge of each crystal, and ratchets the yaw of the crystal with the beam parallel to the inner channel surface by alternating knife-edge scans and rocking scans (up-to-date scripts are available on the I22 GitHub page). After this, the script rotates the crystal into its diffraction condition, and optimizes the Bragg peak through the crystal. This whole procedure takes about 20 minutes for the upstream crystal, and 25 minutes for the downstream crystal, after which the instrument is ready for use. Care is taken in these scripts to ensure that the PIN diode is well aligned with the beam at all stages of the alignment procedure. While the PIN diode offset can be computed from the channel width and the Bragg angle, the penetration depth of the beam is not taken into account in such calculations, and may lead to significant offsets requiring a separate optimization of the PIN diode position. 

For interleaved USAXS/SAXS/WAXS measurements, only the downstream analyzer crystal and PIN diode are moved out of the beam. The upstream crystal as well as the sample are left undisturbed. It must be noted that the direct beam in this case is horizontally offset by 10.8 mm with respect to the usual direct beam due to its travel through the upstream crystal, and that the beamstop on the SAXS detector must be moved accordingly. The upstream crystal does not introduce new artefacts in the beam, but instead cleans up the beam from slit scattering artefacts (in the horizontal plane). 


\section{Performance numbers}


The fine yaw rotation -- used for the scanning motion of the crystal -- is equipped with an encoder strip. When motions of 100 nanoradian are requested, the deviation between the intended positions and the actual positions as reported by the encoder vary by $\pm$ 20 nrad (c.f. Appendix \ref{ax:trackingErrors}), identical to the encoder resolution. As we are aiming to measure a crystal rocking curve with an ideal FWHM of about 7 microradian, an uncertainty of $\pm$ 20 nrad is acceptable. These rotation stages are therefore more than sufficiently capable, if not somewhat overengineered. The full scanning range of the herein presented fine yaw rotation spans about 60 mrad, largely dependent on the range and position of the linear actuator that drives the arm. For practical USAXS scans, no more than $\pm$ 1 mrad is needed (requiring an actuator travel of no more than 1 mm). This also implies that a less expensive, shorter range fine linear actuator may be selected.

The coarse rotation stages (situated on top of the fine rotation stages) allow for the crystal to be moved to various diffraction angles, and placed parallel to the beam for alignment purposes. These stages have a practical resolution approaching 5--10 microradian, which can be sufficient to optimize the upstream crystal rotation: the rotation resolution only needs to be about an order of magnitude better than the divergence of the incident beam. From all X-rays in the incident beam, the channel-cut crystal only selects a 7 microradian FWHM angular segment, i.e. those X-rays of each wavelength in the incident beam that fulfill the Bragg condition within the beam’s divergence. Therefore: 

\begin{itemize}
    \item the upstream rotation stage angular precision only needs to be good enough to be able to pick out a segment of the diverging beam impinging on the crystal surface, and
    \item an intensity loss is observed, proportional to the differences in divergence widths between the incident beam and the Darwin width of the crystal.  
\end{itemize} 
As the USAXS instrument is intended to be interleaved with normal SAXS measurements, the beamline settings for normal experiments are maintained with no effort expended to reduce the divergence of the beam for the USAXS experiments.\\ 

\begin{figure}
	\centering
	\includegraphics[width=\textwidth]{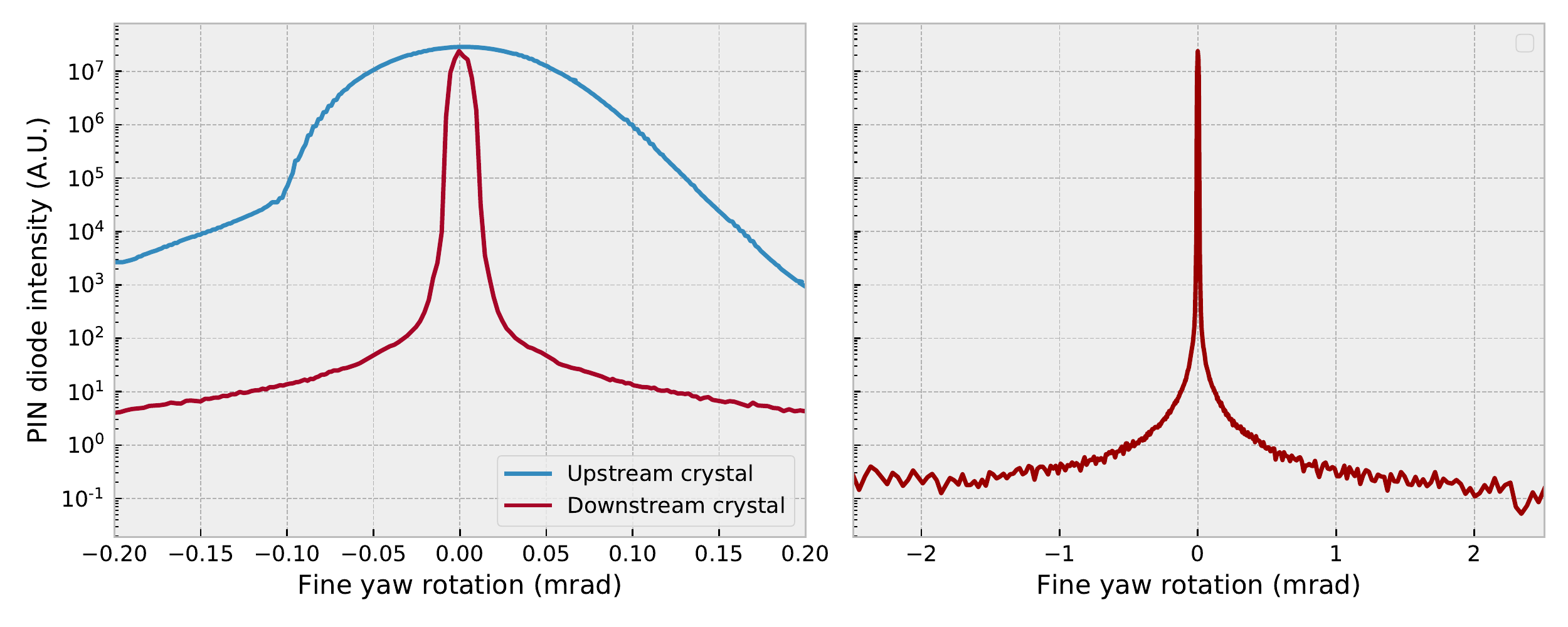}\\[1em]
	\caption{Left: Rocking curves of the crystals on the upstream and downstream rotations, demonstrating that the maximum intensity is not significantly reduced by the second crystal. Right: rocking curve of the downstream crystal over the normal scan range showing the dynamic range available for measurements.} \label{fg:RockingCurves}
\end{figure}

Rocking curves for both the upstream and downstream crystal stages are shown in Figure \ref{fg:RockingCurves}. The rocking curve of the upstream crystal is a convolution of the crystal rocking curve and the divergence inherent in the incident beam. The rocking curve of the downstream crystal, however, matches the divergence of the beam emerging passed through by the upstream crystal. Due to the angular filtering by the upstream crystal, 80\% of the primary beam flux is absorbed by the crystal as it does not match the crystal reflection condition. The downstream crystal, however, passes through most of the remaining flux without large losses. The overall efficiency of the set-up thus reduces the primary beam photon flux by about an order of magnitude. The second rocking curve has a FWHM of approximately $\sqrt{2}$ times the Darwin width expected for a Si(220) reflection at 18 keV, i.e. $\sqrt{2}\times 7$ $\mathrm{\mu}$rad. This measure is the angular selectivity with which the USAXS scan can be performed, and is the first of two critical performance measures. 

The second performance measure is the available dynamic range in the (PIN-diode) detection system. This dynamic range is defined here as the ratio between the maximum and minimum detected intensity, and thereby constrains the samples that can be measured on it; only samples exhibiting a scattering signal measurably larger than the detector background can be considered amenable to this technique. This is also the reason for the multiple crystal reflections in the channel cut, as every reflection suppresses the off-angle signal by another few orders of magnitude. The dynamic range is an interplay of many factors, chief of which are the crystal imperfections, the detector dynamic range, the primary beam intensity, and the quality of background reduction. The latter can be improved through shielding around the crystals and detector, and by starting out with good spectral purity of the primary beam, including a sufficiently effective higher harmonics rejection. As shown from the downstream crystal rocking curve (Figure \ref{fg:RockingCurves}, right-hand side), a dynamic range of about $5\times 10^7$ could be achieved here on a 1 milliradian-wide scan (minimum intensity at the edges of the scan). 


At the sample position, the beam size was scanned using knife edge scans. The beam width was found to be 0.31 mm wide (FWHM) by 0.13 mm high (FWHM), with near-Gaussian profiles. The divergence was estimated by performing a second knife-edge scan 1.3m upstream of the sample, where the vertical beam dimension is approximately 0.15 mm high (FWHM). This corresponds to a vertical divergence of 15 $\mathrm{\mu}$rad.

To minimize mechanically induced variations in the scans, scans are always performed in the same direction, where the linear actuator pushes against the sine arm. This is done to avoid relying on the spring overcoming the stiction (static friction) in the cross-roller bearing. However, other sources of mechanical instability are present, including minor temperature fluctuations, grooving (wear) at the contact point between the linear actuator and the arm contact plate, and friction variations in the bearing. Every scan, therefore, may contain an offset from its previous one (although the encoder ensures a repeatable angular step). To test this, ten subsequent scans were performed for a given sample. These scans show a gradual shift of the $q_0$-value -- the zero-angle position deemed to lie at the center of mass of the transmitted beam -- of approximately 1.3 microradian per scan. By means of a post-processing step, i.e. shifting the collected data to zero around this value, the scans are made to overlap.

With a step-scanning method, the duration of the scan depends on the scan range, the number of datapoints, and the collection time at each datapoint. Practically, this means that a scan for the purpose of extending the SAXS range may require up to 450 datapoints to be collected, requiring about ten minutes per scan. If more overlap between SAXS and USAXS is desired, a wider range must be selected. As an effective speed optimization, scan points were selected on a log-lin-log scale, linearly spaced for the scan over the direct beam, and logarithmically spaced for angles outside that range.

In particular, when PIN diodes are used, such scans can be sped up considerably by using a ``flyscan''-method, as employed at the APS \cite{Ilavsky-2018}. In such a scan, the fine yaw rotation is in continuous motion (at either a constant speed or a more sophisticated speed profile). During this motion, the PIN diode values and the encoder readout are timestamped and read out continuously, and can be rebinned in the post-processing stage. 

\begin{figure}
	\begin{center}
		\includegraphics[width=1\textwidth,angle=0]{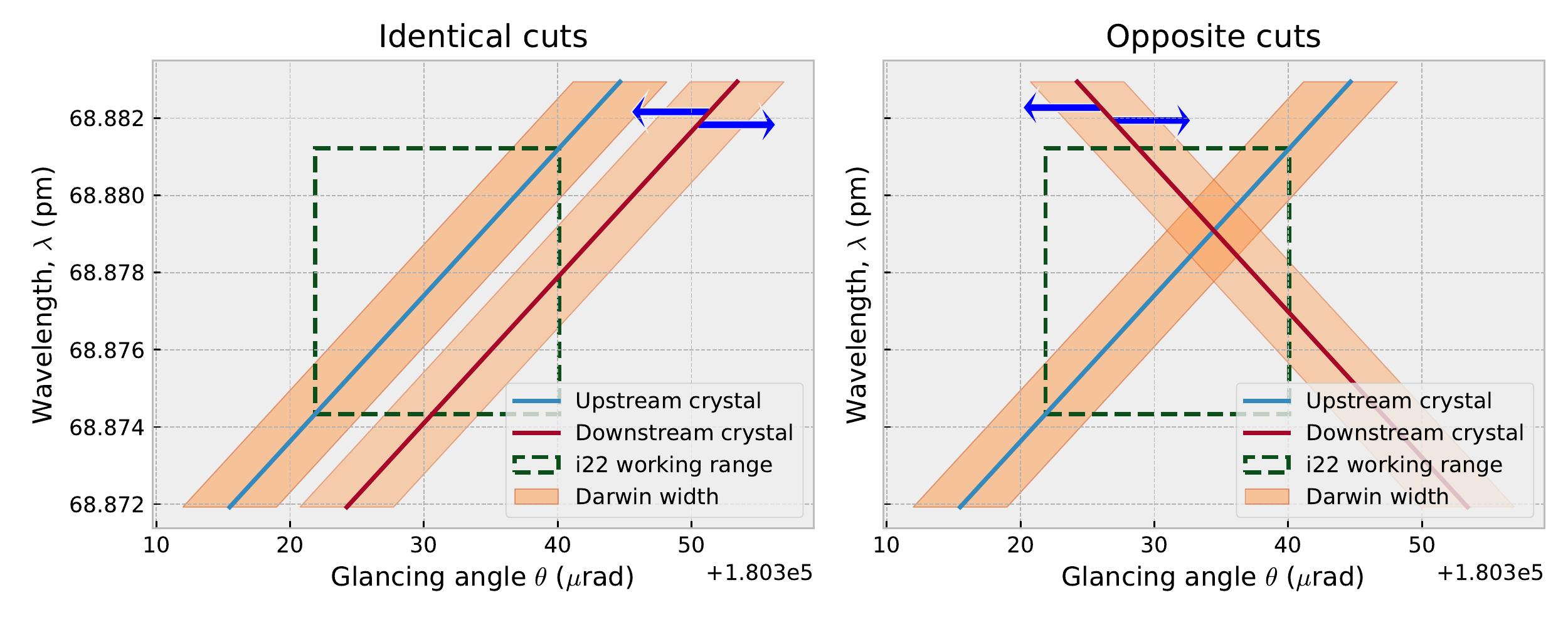}
		\caption{DuMond diagram showing the configuration of the module with parallel-cut crystals (left) and opposite-cut crystals (right). Blue arrows indicate the shift of the pass-through window upon rotation of the crystal. Overlap between the two bands signifies X-rays being reflected through both crystals.} \label{fg:DuMond}
	\end{center}
\end{figure}

To prevent similar misunderstandings in the future, the DuMond diagrams in Figure \ref{fg:DuMond} show why our initial attempt using opposite cut crystals was not successful in establishing a narrow angular selectivity of the downstream crystal \cite{DuMond-1937}. With identical cuts, a rotation of the crystal will pass through radiation only when the two shaded regions overlap, i.e. over a narrow angular range of about 10 $\mu$rad. The radiation that is passed through may consist of a (relatively) wide range of energies. When opposing crystals are chosen instead, in a similar geometry as Bartels monochromators \cite{Bartels-1983}, the diffracted energy range is much more narrow as the overlapping range is now much smaller. However, a decent flux of this narrow energy range will be detected over a much wider angular range, and is, therefore, unsuitable for USAXS. 

\section{Stability and reproducibility}

\begin{figure}
	\begin{center}
		\includegraphics[width=1\textwidth,angle=0]{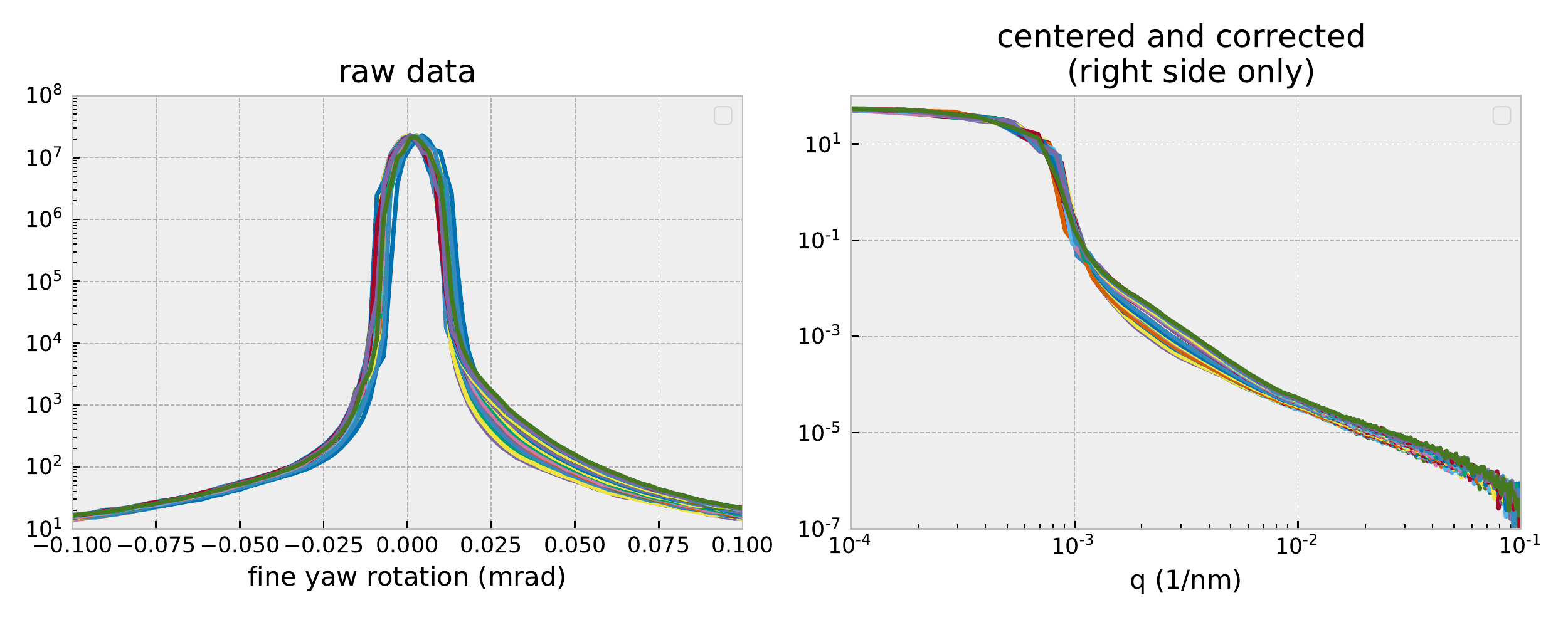}
		\caption{34 USAXS scans of air, interleaved with SAXS/WAXS measurements, to assess the reproducibility of the interleaved operation mode. A gradual change can be observed due to prolonged exposure of the crystals to the X-ray beam, indicating a necessity for regular cleaning of the crystal surfaces.} \label{fg:repeatability}
	\end{center}
\end{figure}

The stability of the complete module with the two towers was tested by comparing multiple scans over the direct beam, while performing interleaved USAXS/SAXS/WAXS measurements. This allows us to test the combined effects of beam stability as well as the upstream- and downstream tower stabilities. The repeatability scans in figure \ref{fg:repeatability} show a very high degree of stability of the module, with only a minor drift visible of the downstream crystal rotation. This is expected, and for analysis the zero-point of every scan is determined anew. The beam intensity shows a minor fluctuation, due to thermal instabilities of $\pm$ 0.15 $^\circ$C in the beamline optics hutch during operation. Additionally, a gradual increase in one segment of the scans, highlighting the need for regular cleaning of the crystal surfaces upon prolonged exposure to high-intensity X-ray beams.

These usability tests demonstrate that the crystal can be moved out and in without suffering notable shifts in the beam. It can occur that the first several moves will shift the apparent beam center through the downstream crystal, as strain on the cables is not yet stabilized. 




\section{Calibration}

To verify the angular resolution, a Si$_3$N$_4$ grating was placed in the sample position. This is a grating consisting of 50 nm bars with a 50 nm air gap, giving an overall period of 100 nm. The bars are held in place by perpendicular supporting ribs spaced approximately 1000 nm apart. In the performance tests, both spacings have been measured. The DAWN powder calibration tool was then used on the resulting data to verify the accuracy of $q$-values obtained from purely geometrical considerations (i.e. actuation arm length and linear actuator motion), and was found to match to 99\%. 



\section{Amenable materials}

\begin{figure}
	\begin{center}
		\includegraphics[width=1\textwidth,angle=0]{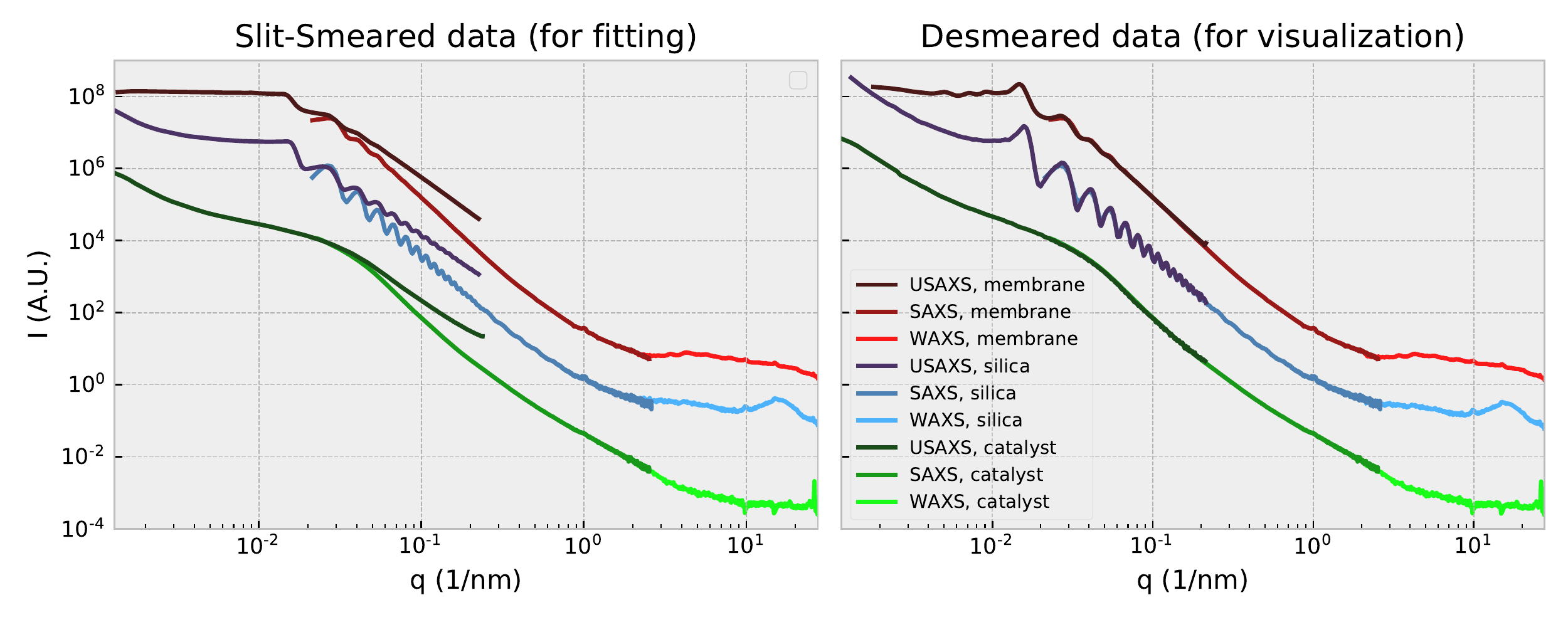}
		\caption{USAXS/SAXS/WAXS patterns of dry 500 nm silica diameter spheres, an alumina membrane and a porous carbon catalyst. Data has been corrected for transmission and background, and is shown in slit-smeared (left), and desmeared form (right)} \label{fg:examples}
	\end{center}
\end{figure}

A range of materials has been subjected to the interleaved measurement procedure. These include aerogels, membranes, porous materials, and powders. Three examples are shown in Figure \ref{fg:examples}, showing the USAXS/SAXS/WAXS scattering patterns of a powder of silica spheres with a diameter of 500 nm, an alumina membrane \cite{Yildirim-2019}, and a porous carbon catalyst \cite{Schnepp-2013}. The left-hand figure shows the data with the USAXS data in its slit-smeared form, with the right-hand figure showing a representation that is more pleasing to the eye: i.e. with desmeared or deconvoluted USAXS data. As the deconvolution method attempts to address a mathematically ill-posed problem, it may introduce or amplify artefacts in the data, and is not recommended for USAXS data analysis. The analysis method least susceptible to misinterpretation involves slit-smearing of the model used in the fitting procedure. 

\begin{figure}
	\begin{center}
		\includegraphics[width=1\textwidth,angle=0]{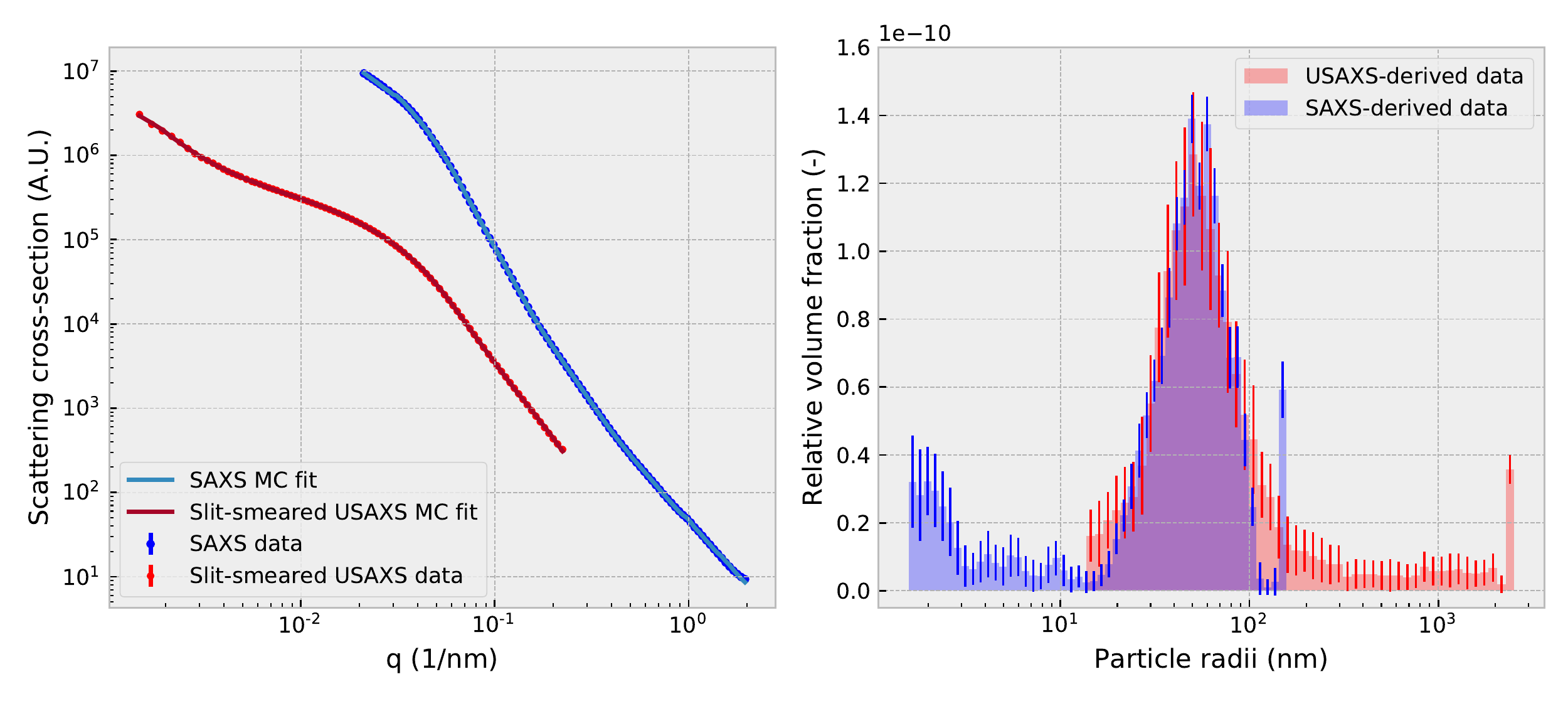}
		\caption{Analysis of the porous carbon catalyst scattering from the SAXS and (slit-smeared) USAXS patterns (left), results in complementary size distributions (right). The bin at the largest size is often abnormally large, effecting a $q\propto I^{-4}$ background slope.} \label{fg:zoe}
	\end{center}
\end{figure}

An example of such an analysis is shown in Figure \ref{fg:zoe}. Here, the SAXS and (slit-smeared) USAXS datasets have been analyzed using the minimal-assumption McSAS analysis method \cite{Bressler-2015}. A prototype implementation of a slit-smearing algorithm has been applied to the MC model during the fitting procedure, to correctly match it to the the slit-smeared USAXS data. 

The distinct advantages of this USAXS module are apparent when paired with a high-performance SAXS instrument that can measure beyond where the USAXS' narrow angular selectivity becomes a disadvantage. At this point, the downstream USAXS crystal can be moved out of the way so that a SAXS measurement can be performed. The downstream crystal can be moved back into position without significant offsets, so that subsequent measurements do not require realignment. Additionally, as the upstream crystal and sample are left untouched, all measurements can be performed at the same location on the sample. 

This procedure was trialled and found to perform well. For the measurements presented here, the USAXS part takes 10 minutes to measure, the SAXS/WAXS part 2 seconds, with approximately 20 seconds required for the downstream crystal and PIN diode to move between configurations.

\section{Outlook} \label{sc:futureImprovements}

As expected, the practical tests revealed a range of possible improvements that may be implemented in future experiments. These are given below. 

\begin{enumerate}
\item{} The motorized pitch and roll rocking stages were found to be unnecessary for the narrow q-range scanned with this USAXS module. While for wider q-ranges the respective alignment of the upstream and downstream crystals needs to be optimized, this is not necessary this close to the direct beam. Therefore, to reduce complexity (and remove four motors), these stages can be replaced by manually adjusted rocking stages, reducing the cost by 1.4 kEuro per crystal tower. 




\item{} The upstream fine-yaw rotation may not be necessary for all instruments, as the coarse-yaw rotation can have sufficient resolution to match the divergence and monochromaticity of the primary beam. That means that the cross-roller bearing, linear actuator, and the encoder can be omitted from the upstream stage, saving a total of 2.7 kEuro. The omissions then either make space for inclusion of a vertical translation, or can be replaced by similarly-sized spacers. 

\item{} The rotations were found to be moderately sensitive to being bumped. The addition of a cowling around the crystal stages may prevent such bumps from interfering with the alignment of the crystals. At the same time, the sample positioning mechanism should be reconsidered and simplified, so that the time spent with the hands near the aligned components is minimized. Currently, a removable filter holder (Thorlabs CFH-1) is used for positioning samples into the beam, but this can be replaced by kinematic magnetically-held sample holders for improved sample handling. 

\item{} The extreme excess of scanning resolution may offer a simplification in the design, in that the sine bars (actuator arm) may be shortened considerably. Alternatively, the current scanning resolution may allow a higher order reflection to be utilized -- such as the Si(440) reflection -- with a much narrower rocking curve, allowing another order of magnitude in $q$ to be gained at the cost of a proportional decrease in intensity.

\item{} A linear actuator with a much smaller motion (of no more than 1 mm) would be sufficient instead of the rather long PI M230.25S. Folded models may be considered, but can be less suited due to their considerably larger backlash.

\item{} The addition of encoders on the coarse yaw rotations as well as on the linear travel stages would simplify the alignment and troubleshooting procedures. A higher reproducibility is also expected if these stages are then controlled with a closed-loop feedback. 

\item{} The PIN-diode used as a detector should be placed on a long-travel ($\geq$ 30 mm) horizontal, and a short-range ($\approx$ 10 mm) vertical stage. This simplifies the alignment of the PIN diode in the beam, and allows its independent motion with respect to the SAXS platform. 

\item{} A significant amount of radiation may pass through the crystal faces at every bounce, particularly for the upstream crystal. An appropriate shield should be placed immediately against or behind the crystal to prevent these unwanted beam(s) from interfering with the measurement.  

\item{} The sample section should be kept as short as feasible, to minimize air scattering as well as the distance to the WAXS detector. 

\item{} There is a lot to be gained from minimizing the time required for scanning and configuration changes. This could include implementing fly-scans, optimizing stage travel speeds, and fine-tuning scan motion profiles. 

\item{} Additional effort to reduce installation time of the equipment will help lower the barrier to use. This could include attaching kinematic reference points to the module, integrating cable management.

\item{} The total cost can be reduced by about 9 kEuro by selecting less expensive motor controllers (such as Trinamic motor drivers) to replace the Omron Delta Tau controllers installed at the Diamond Light Source. Also, 1.4 kEuro can be saved per stage by choosing manual pitch and roll stages. The total cost of a new version could thus be reduced to 30 kEuro.

\item{} The detector can be replaced with an avalanche photodiode, which have a higher dynamic range. 

\end{enumerate}

\section{Conclusions}

The USAXS module is a commendable addition to existing high-performance SAXS instruments, so that their $q$ range may be extended by another decade. The instrument has proven itself to be a low-cost addition, stable enough to be shifted in for interleaving USAXS measurements with SAXS/WAXS measurements. By restricting its measurement range to the ultra-small angles at which it performs most efficiently, two additional benefits are secured. Firstly, the infinite-width slit smearing assumption holds at these small angles with the chosen beam size and detector entrance aperture. Secondly, the results are less sensitive to misorientation of the crystal planes of the two channel-cut crystals with respect to each other. 
The USAXS module has been demonstrated to be useful for a range of practical materials, and the interleaved SAXS/USAXS experiments show that its concept is sound. Future improvements are expected to further simplify and speed up the installation, alignment and measurement procedures. 

The designs of the instrument components are available under a CC-BY license. 



\appendix

\section{Bill of materials}

%

\begin{table}
\caption{Bill of Materials for a single crystal stage. \label{tb:costUDStage}}
\begin{tabular}{L{0.3\textwidth}L{0.3\textwidth}R{0.1\textwidth}R{0.1\textwidth}R{0.1\textwidth}}      
Item & Variant & amount & Price per (kEuro, ex VAT) & Total price \\
\hline
Coarse yaw rotation & Kohzu RA07A-W01 with 2-phase stepper motor & 1 & 1.93 & 1.93\\ 
Linear rail profile & QIOptik (LINOS) FLS 95-500-M & 1 & 0.11 & 0.11\\ 
rail clamp for tower & Thorlabs XT95P11/M & 1 & 0.072 & 0.072\\ 
Cross-roller rotation bearing & THK RU66 UU CC0 USP & 1 & 0.79 & 0.79\\ 
Linear actuator for rotation & PI M230.25S (25 mm linear actuator) & 1 & 1.28 & 1.28\\ 
Interferometer strip & Heidenhain LIDA 489x70mm & 1 & 0.09 & 0.09\\ 
Interferometer head & Heidenhain LIDA 48 & 1 & 0.529 & 0.529\\ 
Tower horizontal translation & Newport UTS 100 PP & 1 & 2.782 & 2.782\\ 
Pitch and roll rotations & Kohzu SA05B-RS01 & 1 & 2.7 & 2.7\\ 
motor cables & Kohzu CB03 & 3 & 0.05 & 0.15\\ 
Optical breadboard & MB4515/M & 1 & 0.115 & 0.115\\ 
Channel-cut crystal Si(220) & (custom manufactured) & 1 & 2.5 & 2.5\\ 
&  &  &  & \\ 
SUBTOTAL: & - & - & - & 13.048\\ 
\end{tabular}
\end{table}

\begin{table}
\caption{Bill of Materials for the central sample stage.\label{tb:costCStage}}
\begin{tabular}{L{0.3\textwidth}L{0.3\textwidth}cR{0.1\textwidth}R{0.1\textwidth}}      
Item & Variant & amount & Price per (kEuro, ex VAT) & Total price \\
\hline
Optical breadboard & MB4515/M & 1 & 0.115 & 0.115 \\
Horizontal translation & Newport UTS 100 PP & 1 & 2.782 & 2.782 \\
Vertical translation & Newport UTS 100 PP & 1 & 2.782 & 2.782 \\
&  &  &  & \\ 
SUBTOTAL: & - & - & - & 5.679\\ 
\end{tabular}
\end{table}

\begin{table}
\caption{Bill of Materials for associated components.\label{tb:costAssComponents}}
\begin{tabular}{L{0.3\textwidth}L{0.3\textwidth}cR{0.1\textwidth}R{0.1\textwidth}}      
Item & Variant & amount & Price per (kEuro, ex VAT) & Total price \\
\hline
Optical base & Thorlabs MB4545/M & 1 & 0.25 & 0.25 \\ 
Low-profile screws & Thorlabs SH6M10LP & 2 & 0.021 & 0.042 \\ 
Motor controllers & Trinamic TMCM6110  & 2 & 0.2 & 0.4 \\ 
or:  & Omron Delta Tau Geobrick LV IMS II (8 motors) & 1 & 6.253 & 6.253 \\ 
 & Omron Delta Tau Geobrick Motor Power supply (8 motors) & 1 & 2.488 & 2.488 \\ 
 & Renishaw Tonic Interpolator 1000x & 2 & 0.369 & 0.738 \\ 
PIN diode & Hamamatsu S3590-09 & 1 & 0.2 & 0.2 \\ 
Diode Amplifier & FEMTO DLPCA-S2 & 1 & 2.5 & 2.5 \\ 
 &  &  &  &  \\ 
SUBTOTAL:  & - & - & - & 12.579 \\ 
\end{tabular}
\end{table}

\section{Fine yaw motion}\label{ax:trackingErrors}

\begin{figure}
	\begin{center}
		\includegraphics[width=1\textwidth,angle=0]{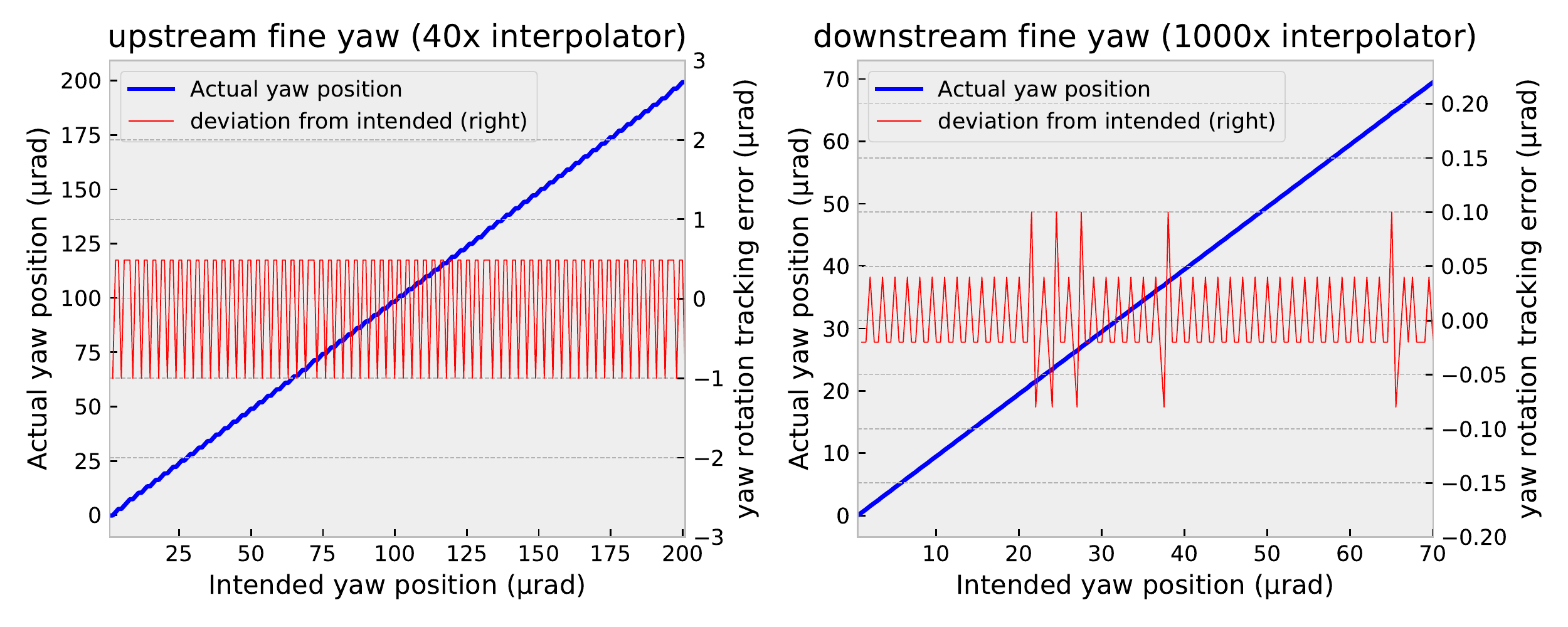}
		\caption{Intended versus actual positions for the upstream and downstream fine-yaw motions over a finely-stepped scan. Upstream tracking errors are larger due to the reduced precision of the interferometer interpolator (40x) versus that of the downstream stage (1000x) and are not expected to be indicative of the actual stage positioning accuracy.} \label{fg:tracking}
	\end{center}
\end{figure}

The fine yaw motion for both the upstream and downstream stages (with 40x and 1000x interpolator on the interferometer strip, respectively), has been assessed for positioning reliability. This was done by comparing the final resting position as read out on the encoder, with the intended position that the motor was driven to. The resulting tracking errors are shown in Figure \ref{fg:tracking}

\ack{Acknowledgements}

The authors thank Laura Forster for her active interest in the experiments and her participation in modeling for the flight case photograph. 

\bibliographystyle{iucr}
\referencelist[InexpensiveUSAXS.bib]

\end{document}